\documentclass{article}
\usepackage{spconfa4,amsmath,graphicx,hyperref, amssymb,amsfonts}
\usepackage{algorithm2e}
\usepackage{svg}
\usepackage{enumitem} % mezerování v odrážkách
  \setlist{topsep=0pt,partopsep=0pt,noitemsep}

\DeclareMathOperator*{\argmin}{arg\,min}
\def\x{{\mathbf x}}
\def\u{{\mathbf u}}
\def\y{{\mathbf y}}
\def\p{{\mathbf p}}
\def\q{{\mathbf q}}

\title{Audio dequantization using instantaneous frequency}
\name{Vojtěch Kovanda, Pavel Rajmic\thanks{The work was supported by the Czech Science Foundation (GAČR) Project No.\,23-07294S.}}
\address{\textit{Dept.\ of Telecommunications} \\
\textit{Brno University of Technology}\\
Czech Republic\\
xkovan07@vutbr.cz, pavel.rajmic@vut.cz}
\begin{document}
%\ninept
%
\maketitle
\begin{abstract}
We present a dequantization method that employs a phase-aware regularizer, originally successfully applied in an audio inpainting problem. The method promotes a temporal continuity of sinusoidal components in the time-frequency representation of the audio signal, and avoids energy loss artifacts commonly encountered with $\ell_1$-based regularization approaches. The proposed method is called the Phase-Aware Audio Dequantizer (PHADQ). The method is evaluated against the state-of-the-art using the SDR and PEMO-Q ODG objective metrics, and a~subjective MUSHRA-like test.
\end{abstract}

\section{Introduction}
Quantization is a nonlinear distortion that arises from rounding during the A/D conversion or bit-depth reduction.
This distortion appears as the quantization noise, which, at low bit depths (less than 16 bits per audio sample), results in unpleasant auditory artifacts \cite{Watkinson2001:Art.of.Digital.Audio}.
Dequantization is the attempt to recover a~signal close to its original form, based on the observed quantized samples.
Such a problem is ill-conditioned; consequently, the reconstruction process requires some form of regularization.
For example, exploiting sparsity \cite{ZaviskaRajmicMokry2021:Audio.dequantization.ICASSP, BrauerGerkmannLorenz2016:Sparse.reconstruction.of.quantized.speech.signals} is a commonly used strategy. 
% Sparsity is commonly used as a regularizer in related problems, such as inpainting and declipping.
Autoregressive modeling represents another line of approach to dequantization \cite{TroughtonGodsill1999:MCMC.restoration.quantised.time.series,Troughton1999:Dequantization.Sinusoidal.AR,TroughtonGodsill2001:MCMC.restoration.nonlinearly.distorted.AR.signals}.

In this paper, we present a dequantization method that employs a phase-aware regularizer introduced in \cite{TanakaYatabeOikawa2024:PHAIN}, originally successfully applied in the context of the audio inpainting problem.
The regularizer maintains the temporal continuity of sinusoidal components in the audio signal time-frequency representation and avoids the energy loss artifacts commonly encountered with $\ell_1$-based regularization approaches
\cite{ZaviskaRajmicMokry2021:Audio.dequantization.ICASSP,MokryRajmic2020:Inpainting.revisited}.
The goal of this paper is to investigate whether the phase-aware regularization is beneficial also in the dequantization setup.
The reconstruction quality and computational demands are compared to a~baseline audio dequantization method~\cite{ZaviskaRajmicMokry2021:Audio.dequantization.ICASSP}.

\section[Review of PHAIN (Phase-aware audio inpainter)]{Review of PHAIN\\ (Phase-aware audio inpainter)}
Audio inpainting is the task of recovering portions of an audio signal that are considered lost or unreliable \cite{Adler2012:Audio.inpainting,Mokry2022:Audio.inpainting.NMF,MokryRajmic2025:Inpainting.AR,MokryRajmic2020:Inpainting.revisited}.
Let $\y \in \mathbb{R}^L$ be an audio signal with unreliable parts.
PHAIN~\cite{TanakaYatabeOikawa2024:PHAIN}
% (Phase-Aware Audio Inpainter)
aims to reconstruct 
$\y$ using the instantaneous frequency of the signal by solving the following optimization problem:
\begin{equation}
    \label{PHAIN.eq}
    \hat{\x} = \argmin\limits_{\x\in\mathbb{R}^L} \ \ \lambda\| D R_{\omega_{\mathbf{s}}} G_g \x\|_1 + \imath_{\mathit{\Gamma}}(\x) .
\end{equation}
Problem \eqref{PHAIN.eq} contains a penalty 
$\| D R_{\omega_{\mathbf{s}}} G_g \x\|_1$  weighted by a~positive $\lambda$, and an indicator function $\imath_{\mathit{\Gamma}}(\x)$ taking value $\infty$ when $\x \notin \Gamma$,
thus enforcing the estimation to retain the reliable parts of the signal corresponding to the set of feasible signals~${\mathit{\Gamma}}$.
The linear operators in the penalty function are:
\begin{itemize}
    \item
    The discrete Gabor transform (DGT) $G_g$ employing a~window~$g$.
    \item 
    The phase correction $R$ of a spectrogram, based on the time derivative of the phase $\omega_\mathbf{s}$ (i.e., the instantaneous frequency) of a~signal $\mathbf{s}$.
    Phase correction is performed using the cumulative sum of $\omega_\mathbf{s}$ as follows \cite{YatabeiPCTV}:
    \begin{equation}
    \label{correction.eq}
    (R_{\omega_\mathbf{s}} \mathbf{z})[m,n] =  \text{e}^{-2 \pi \text{i} a \sum_{t=0}^{n-1}\omega_\mathbf{s}[m, t]/M}z[m,n],
    \end{equation}
    % \eqref{correction.eq}.
    where $a$ denotes the hop-size of the Gabor transform, and $n,m$ represent the time and frequency indices, respectively.
    Size of $\omega_\mathbf{s}$ is identical to the size of the spectrogram obtained by $G_g$.
    \item 
    The time-directional first-order difference \begin{equation}
    \label{tdvariation.eq}
    (D \mathbf{z})[m,n] =  z[m,n]-z[m,n+1].
    \end{equation}
\end{itemize}
The convex $\ell_1$-norm denoted $\|\cdot\|_1$ quantifies the time-directional variation of the phase-corrected time-frequency representation of $\x$.
When used as a~penalty, it promotes sinusoidal components and, conversely, attenuates components whose phase cannot be adequately approximated as a~linear function, such as noise or transients.
As a consequence, PHAIN to a~great extent solves the problem encountered in plain sparsity-based methods which is loosing signal energy inside the gap
\cite{MokryRajmic2020:Inpainting.revisited}.
When $\omega_\mathbf{s}$ is fixed, the problem
\eqref{PHAIN.eq}
is convex and solvable via the Chambolle--Pock algorithm \cite{ChambollePock2011:First-Order.Primal-Dual.Algorithm,Condat2023:Proximal.splitting.algorithms}.

Based on the regime of treating the instantaneous frequency during the process of reconstruction, PHAIN can further be classified as follows:
\begin{itemize}
    \item B-PHAIN, which uses the instantaneous frequency $\omega_\mathbf{y}$ of the degraded signal, i.e.\ including unreliable parts.
    \item U-PHAIN, which updates the instantaneous frequency
    % $\omega_\mathbf{y}$
    according to the estimate of $\hat{\x}$ during iterations.
\end{itemize}

\section{Method}
We call the proposed method simply the Phase-Aware Audio Dequantizer (PHADQ).
We employ the same penalty as in PHAIN, but for a~different restoration task.
To further explain motivation, we provide supplementary figures\footnote{\url{https://vojtechkovanda.github.io/PHADQ/}}.

Let $\y^\text{q} \in \mathbb{R}^L$ be a quantized observation of an original audio signal with a~quantization step $\Delta$. When employing uniform quantization, the quantization step size is fixed and given by $\Delta = 2^{-w+1}$, where $w$ denotes the word length in bits per sample (bps). We recover the signal by solving the same convex problem as in \eqref{PHAIN.eq}, except that now no reliable parts of the signal are available.
Therefore, in PHADQ, the set of feasible signals is given as:
\begin{equation}
    \label{Gama.eq}
    \mathit\Gamma = \{\y \in \mathbb{R}^L \ | \ \|\y-\y^\text{q}\|_\infty < \Delta/2\},
\end{equation}
which means that a~feasible sample must lie within the quantization interval corresponding to the original signal.

Using the indicator function $\imath_{\mathit{\Gamma}}$, we obtain the so-called consistent variant of the problem.
An alternative approach is to relax the constraint defined by the indicator function and employ the (euclidean) distance to the set of feasible samples $d_\Gamma$. Problem \eqref{PHAIN.eq} is then transformed to its inconsistent variant:
\begin{equation}
    \label{incons.eq}
    \hat{\x} = \argmin\limits_{\x\in\mathbb{R}^L} \ \ \lambda\| D R_{\omega_{\mathbf{s}}} G_g \x\|_1 + \cfrac{1}{2}\,d^2_\mathit\Gamma(\x) .
\end{equation}
We differentiate between B-PHADQ and U-PHADQ based on whether the instantaneous frequency estimate is updated or kept fixed during the iterative computation.

\subsection{B-PHADQ}
Both the consistent and inconsistent restoration problems (i.e., \eqref{PHAIN.eq} and \eqref{incons.eq}, respectively) are solved by Algorithm \ref{alg:CP}, which constitutes a specific instance of the Chambolle–Pock algorithm, wherein $\x \in \mathbb{R}^L$, $\p \in \mathbb{R}^L$ is a primal variable, $\q \in \mathbb{C}^P$ is a dual variable and $\u \in \mathbb{R}^L$ is a temporary variable.
Positive scalars $\tau$, $\sigma$ are the step sizes and $\rho$ is a~relaxation factor~\cite{Condat2023:Proximal.splitting.algorithms}. The scalar $\lambda >0$ is the weighting factor for the penalty function, having also an effect on the reconstruction quality and the convergence speed. Before applying the algorithm,
an initial estimate of the instantaneous frequency $\omega_\mathbf{s}$ has to be established, which
% in the case of B-PHADQ
is achieved by the following formula using the quantized observation
% $\y^\text{q}$,
% following
\cite{TanakaYatabeOikawa2024:PHAIN}:
\begin{equation}
    \label{instfreq.eq}
    \omega_{\mathbf{s}}[m, n] =
 \omega_{\mathbf{y}^\text{q}}[m, n] = - \text{Im} \Bigg[\cfrac{(G_{g'} {\mathbf{y}^\text{q}})[m,n]}{(G_g {\mathbf{y}^\text{q}})[m,n]} \Bigg ].
\end{equation}
Here, the Gabor transform is used twice, once with window function $g$ and the second time with $g'$\!,  the time derivative of~$g$.
$\text{Im} [\cdot ]$ denotes the imaginary part.
The initial values $\x^{(0)}\!, \p^{(0)}\!, \q^{(0)}$ are zeros.

\RestyleAlgo{ruled}

\begin{algorithm}[t]%[!h]
    \DontPrintSemicolon
    \caption{B-PHADQ}
    \label{alg:CP}
    Choose parameters
    $\tau, \sigma, > 0, \ \rho \in [0, 1]$ 
    and initial values $\x^{(0)}, \p^{(0)}, \q^{(0)}$;
    initialize $\omega_\mathbf{s}$  by \eqref{instfreq.eq}
    \\
    \For{$i=0,1,\dots$}{
    $\q^{(i+1)}=\text{clip}_\lambda(\q^{(i)}+\sigma DR_{\omega_\mathbf{s}}G_g\x^{(i)})$
    \;
    $\u = \p^{(i)}-\tau G^*_gR^*_{\omega_\mathbf{s}}D^*\q^{(i+1)}$ ~\,\%\,auxiliary
    \;
    $\p^{(i+1)} = \text{proj}_\mathit\Gamma(\u)$ ~\,\%\,consistent
    \;
    $\p^{(i+1)} = \frac{1}{\tau+1}(\tau \ \text{proj}_\mathit\Gamma(\u)+\u)$ ~\,\%\,inconsistent
    \;
    $\x^{(i+1)} = \p^{(i+1)}+\rho(\p^{(i+1)}-\p^{(i)})$
     }
     \Return{$\p^{(i+1)}$}
\end{algorithm}

The algorithm utilizes proximal operators of the  functions present in the minimization, which in the case of the $\ell_1$ norm correspond to the soft-thresholding, and in the case of the indicator function (consistent case) it is the projection onto the set $\mathit\Gamma$.
In the inconsistent case, the proximal operator is \cite{Beck2017:First.Order.Methods}
\begin{equation}
    \label{proxdistance.eq}
 \text{prox}_{\alpha d^2_\mathit\Gamma/2}(\x) = \cfrac{1}{1+\alpha}\big(\alpha\,\text{proj}_\mathit\Gamma (\x) + \x\big).
\end{equation}
The asterisk in Algorithm \ref{alg:CP} denotes the adjoint of a linear operator. %; the adjoint of $G_g$ is the inverse Gabor transform, transforming the signal from the time-frequency domain $\mathbb{C}^P$ back into the time-domain $\mathbb{R}^L$ \cite{Grochenig2001:Foundations.T-F.analysis}. The adjoint of $R$ and $D$ is obtained as follows:
%\begin{equation}
%    \label{adjointD.eq}
%(D^* \mathbf{z})[m, n] = \begin{cases}
%    \ \ \ \ \ \ \ \ \ \ \ z[m,n] \ \ \ \ \ \ \ \ \ \ \ \ \ \ \ \ \ \ \ \ \ \ (n = 0) \\
%    z[m,n] - z[m,n-1] \ \ \ \ \ \ \text{(otherwise)} \\
%  \ \ \ \ -z[m,n-1] \ \ \ \ \ \ \ \ \ \ \ \ \ \ \ (n = N-1),
%\end{cases}
%
%\end{equation}
%\begin{equation}
%    \label{adjointR.eq}
%(R^*_{\omega_\mathbf{s}} \mathbf{z})[m, n] = \text{e}^{2 \pi \text{i} a \sum_{t=0}^{n-1}\omega_\mathbf{s}[m, t]/M}z[m,n].\textbf{}
%\end{equation}
Note that
% In the algorithm \ref{alg:CP},
the term $\p^{(i+1)}$ is computed differently for the consistent and for the inconsistent variant.
Finally, $\text{clip}_\lambda$ is defined as
$\text{clip}_\lambda (\x) = \x-\text{soft}_\lambda(\x)$.

\subsection{U-PHADQ}
As in U-PHAIN \cite{TanakaYatabeOikawa2024:PHAIN}, an alternative approach to problems \eqref{PHAIN.eq} and \eqref{incons.eq} is to regularly update the instantaneous frequency that is fixed in B-PHADQ in Algorithm \ref{alg:CP}.
In U-PHADQ, such an update is performed after
a~user-prescribed number of iterations of Algorithm \ref{alg:CP}.
The new instantaneous frequency is calculated for $\p$ rather than from $\mathbf{y}^\text{q}$,
this way corresponding to the current state of reconstruction. U-PHADQ therefore contains the main loop of Algorithm~\ref{alg:CP} as its inner loop.
% , the same way as in U-PHAIN \cite{TanakaYatabeOikawa2024:PHAIN}.

\section{Experiment and Results}
For the experiment, we selected musical recordings from the IRMAS%
\footnote{\url{https://www.upf.edu/web/mtg/irmas}}
and EBU SQAM%
\footnote{\url{https://tech.ebu.ch/publications/sqamcd}}
datasets, and these selections are evaluated separately.
The same datasets were used to evaluate the performance of PHAIN for audio inpainting~\cite{TanakaYatabeOikawa2024:PHAIN}.

From the IRMAS database, we selected 50 excerpts from various genres, featuring mixtures of different musical instruments (guitar, piano, violin, and vocals) included in the IRMAS-TestingData. The recordings have a sampling frequency of 44.1 kHz. Since they are originally in stereo, only the first channel was used, and each excerpt was truncated to 7 seconds.
From the EBU SQAM database, we selected 10~recordings, mostly of solo instruments. The recordings have a sampling frequency of 44.1 kHz, and segments of approximately 6 seconds were used.

Audio has been peak-normalized before any processing to make the most of the available dynamic range.
The audio signals were quantized to seven different word lengths, ranging from 2 to 8\,bps using the mid-riser uniform quantization \cite{ZaviskaRajmicMokry2021:Audio.dequantization.ICASSP}.
In the experiment, the Gabor transform with a~Hann window of 8,192 samples in length, 75\% overlap and 16,384 frequency channels has been used.
All the computation was performed in \mbox{Matlab} R2025a.

The parameters of Algorithm \ref{alg:CP} were manually tuned. We used $\tau = 1,\ \sigma = 1$ and $\rho = 1/3$.
For the best possible results, the balancing parameter $\lambda$ of the task needs to be set depending on the word length and variant; we set it  as shown in Table~\ref{tab:lambda.wrt.bps},
in which the indices c or i denote consistent or inconsistent variant.
Note that $\lambda$ increases with decreasing bps, which is expected since the lower the bps, the greater the amount of \mbox{error/noise}.
Also note that the inconsistent case requires a~lower $\lambda$ than its consistent counterpart, which makes sense,
because in \eqref{incons.eq}, the error is partially absorbed by the distance term, opposite to the hard constraint in \eqref{PHAIN.eq}.
\begin{table}[h!]
\vspace{-2ex}
\caption{Values of $\lambda$ according to word length and method}
\label{tab:lambda.wrt.bps}
\centering
\setlength{\tabcolsep}{3.5pt}
\renewcommand{\arraystretch}{1.1}
\begin{tabular}{|c||c|c|c|c|c|c|c|c|} % c = center, | = vertical line
\hline
bps & 2 & 3 & 4 & 5 & 6 & 7 & 8 \\ \hline
$\lambda_\text{c}$ & $0.07$ & $0.070$ & $0.030$ & $0.010$ & $0.0010$ & $0.0050$ & $0.0002$ \\ \hline
$\lambda_\text{i}$ & $0.07$ & $0.015$ & $0.006$ & $0.001$ & $0.0008$ & $0.0005$ & $0.0002$ \\ \hline
\end{tabular}
\end{table}

In the experiments, method performance was assessed using the objective metric SDR, which denotes the signal-to-distortion ratio and quantifies the physical similarity between waveforms.
We also employed the PEMO-Q objective difference grade (ODG) \cite{Huber:2006a}, which considers the perceptual characteristics of the audio signals. The ODG is measured on a~scale from $-4$ to 0 (worst to best).
To obtain subjective quality assessments, we employ the MUSHRA listening test~\cite{ITU-R2015:MUSHRA}.

As a baseline for comparison, we chose a sparsity-based dequantization method, namely the  Chambolle--Pock algorithm (CP) performing best in the study \cite{ZaviskaRajmicMokry2021:Audio.dequantization.ICASSP}.
This method solves the optimization  problem
% \eqref{CPsparse.eq} based on sparsity.
\begin{equation}
    \label{CPsparse.eq}
    \hat{\x} = \argmin\limits_{\x\in\mathbb{R}^L} \ \ \lambda\| G_g \x\|_1 + \imath_{\mathit{\Gamma}}(\x),
\end{equation}
seeking for an analysis-sparse time-domain signal which fits the dequantization constraints.
We used the Matlab code from \cite{ZaviskaRajmicMokry2021:Audio.dequantization.ICASSP} with the same DGT settings as in the PHADQ experiment to ensure a fair comparison between the methods.

\begin{figure}[t]%[!h]
    \centering
    \vspace{-2ex}
    \includegraphics[width=\columnwidth]{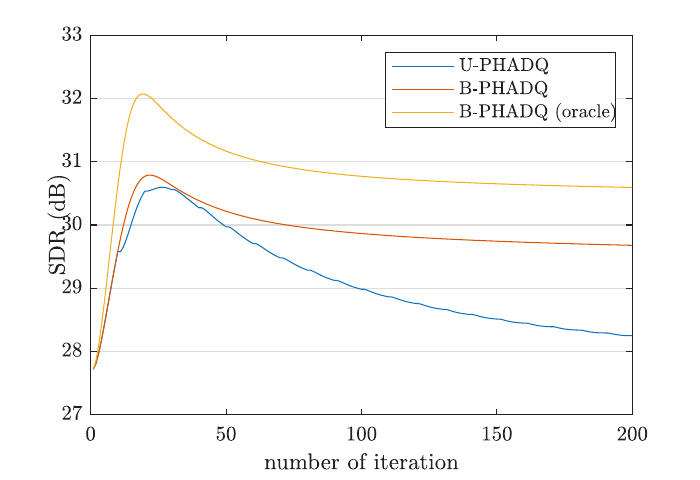}%
    \vspace{-1ex}%
    \caption{Reconstruction quality of a~violin recording (6\,bps word length) across iterations in terms of SDR.
    For \mbox{U-PHADQ}, $\omega_{\mathbf{s}}$ is updated every 10 iterations.}
    \label{fig:iterations:sdr}
\end{figure}

We show an example showing the evolution of SDR across iterations of PHADQ variants presented in Fig.~\ref{fig:iterations:sdr}. When comparing U-PHADQ and B-PHADQ across the EBU SQAM dataset, it was found that updating the instantaneous frequency during iterations (U-PHADQ) does not improve the SDR.
For this reason, the experiments focus only on the B-PHADQ variant. In addition, to see the maximum achievable performance of PHADQ, there is the B-PHADQ (oracle) variant, based on the instantaneous frequency of the original signal, which however cannot be accessed in practical scenarios. For the oracle variant the consistent B-PHADQ was used. 

The results of the PHADQ method for both datasets show ODG and SDR after 60 iterations of the algorithm, which represents a balance between the average number of iterations required to reach optimal SDR and ODG values. We note that a higher number of algorithm iterations does not guarantee improved ODG values; supplementary graphs are available\footnote{\url{https://vojtechkovanda.github.io/PHADQ/}}.
The CP results were obtained after 500 iterations of the algorithm, achieving optimal ODG values \cite{ZaviskaRajmicMokry2021:Audio.dequantization.ICASSP}.
% pozn. více iterací neznamená lepší ODG
%
\subsection{Objective evaluation}
\begin{figure}[!h]
    \centering
    \vspace{-2ex}
    \includegraphics[scale=0.8]{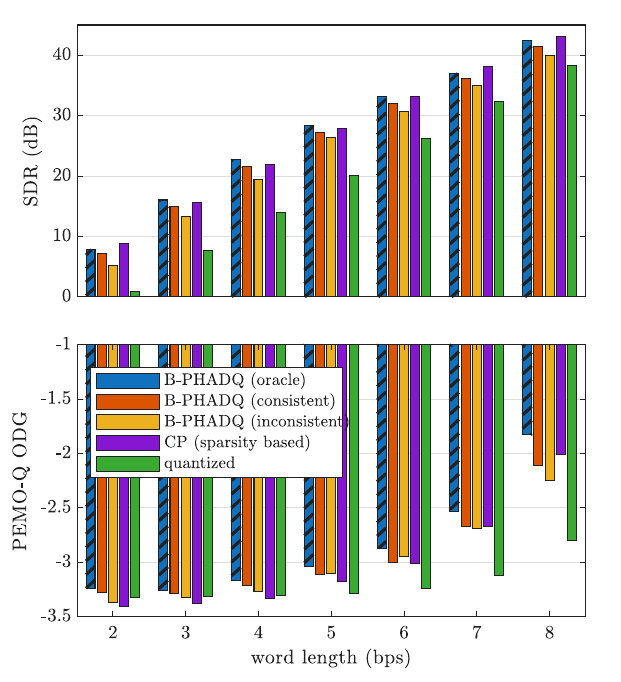}%
    \vspace{-1.4ex}%
    \caption{Average SDR and ODG on the EBU SQAM dataset}
    \label{fig:sqam}
\end{figure}

\begin{figure}[!h]
   \centering
   \vspace{-2ex}
   \includegraphics[scale=0.8]{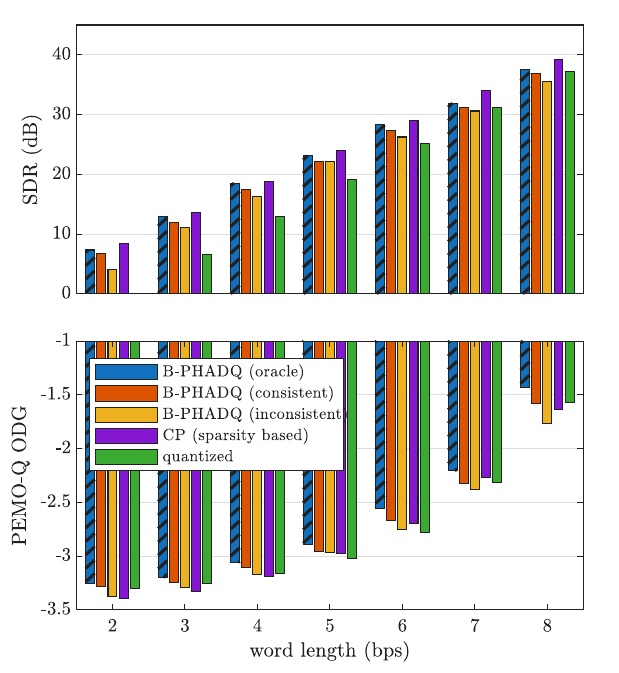}%
   \vspace{-1.4ex}%
   \caption{Average SDR and ODG results on the IRMAS dataset}
   \label{fig:irmas}
\end{figure}

The SDR and ODG results for the EBU SQAM and IRMAS dataset are shown in Fig.~\ref{fig:sqam} and Fig.~\ref{fig:irmas} respectively.
As expected, the reference B-PHADQ (oracle) exhibits the best results in ODG in both datasets.
CP outperforms the proposed methods in terms of SDR. However, in terms of ODG, the consistent B-PHADQ outperforms CP across almost all word lengths in both datasets. In the EBU SQAM dataset, all methods achieve higher SDR but lower ODG. This is due to the fact that EBU SQAM contains recordings of solo instruments, making the differences between recordings more perceptible. On the other hand, the IRMAS dataset contains recordings with multiple instruments, which masks the quantization noise more effectively, resulting in a~higher ODG. %word lengths.
%In terms of ODG, PHADQ proved to be superior, especially at a word length of 8 bps, where both consistent and inconsistent B-PHADQ exceeded CP by more than 0.3.
%Fig.~\ref{fig:iterations} depicts the iteration number resulting in the best SDR and ODG reconstruction. At a word length of 8\,bps, both B-PHADQ variants required no more than 60 algorithm iterations on average.

%The results of the EBU SQAM dataset are presented Fig.~\ref{fig:sqam} in the same way.
%There are some differences compared to the IRMAS dataset. The overall SDR is higher across all word lengths, whereas ODG shows lower values. This effect is due to the fact that the EBU SQAM dataset contains solo instruments, in which the (de)quantization error is hardly masked. In terms of SDR, CP again proves to be better than both consistent and inconsistent B-PHADQ. In terms of ODG, the B-PHADQ variants perform better than CP for most word~lengths; however, the difference is not that pronounced than in the case of the IRMAS dataset.

%\begin{figure}[!h]
%    \centering
%     \vspace{-2ex}
%    \includegraphics[scale=0.8]{IRMAS_iterations.pdf}%
%    \vspace{-1.4ex}%
%    \caption{Average number of iterations achieving the best SDR and ODG on the IRMAS dataset.}
%    \label{fig:iterations}
%\end{figure}

\subsection{Subjective evaluation}
\begin{figure}[t]%[!h]
    \centering
    \vspace{-2ex}
    \includegraphics[scale=0.77]{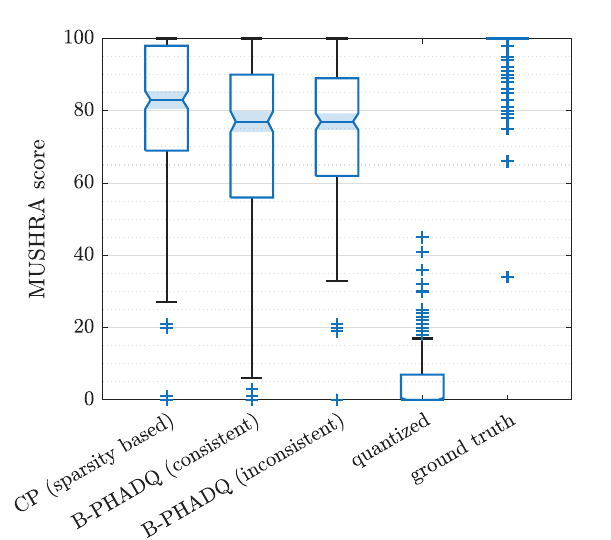}%
    \vspace{-1.4ex}%
    \caption{MUSHRA results for 6 and 7 bps quantization on EBU SQAM dataset}
    \label{fig:sqam:mushra}
\end{figure}

The subjective test was conducted on the EBU SQAM database, comparing the CP, consistent B-PHADQ, and inconsistent B-PHADQ methods. Only 6-bit and 7-bit word lengths were used in the test, as these are the highest bit depths at which quantization distortion is still relatively clearly perceptible. The test results are shown in Fig.~\ref{fig:sqam:mushra} and they are in a~close agreement with the objective evaluation.

\subsection{Computational considerations}
For a 6-second-long excerpt, a single iteration of the consistent B-PHADQ variant
takes about 0.07 second on a PC with AMD 3.80\,GHz CPU and 32\,GB RAM. Therefore a full, 60~ite\-ra\-tion~long reconstruction is obtained in 4.2 seconds. A~single iteration of the baseline sparsity-based CP algorithm takes about 0.03 second, thus a~full reconstruction with 500 iterations requires approximately 15~seconds.
The proposed method thus requires significantly lower computational time to achieve results of comparable  quality.

\section{Conclusion}
In this paper, we introduced an audio dequantization method that employs instantaneous frequency as a regularizer. The proposed approach demonstrated strong performance in objective evaluations.
Moreover, the method requires fewer algorithmic iterations to converge than the sparsity-based CP algorithm, leading to reduced computational time. These findings highlight the potential of the proposed framework for efficient and high-quality restoration of quantized signals.
% zhodnotit poslechové testy

Codes for Matlab are publicly available.\!%
\footnote{\url{https://github.com/vojtechkovanda/PHADQ}}%

\newpage\newpage\newpage
\bibliographystyle{IEEEbib}
\bibliography{literatura}

\end{document}